\documentclass[11pt]{article}
\usepackage{jheppub}
\pdfoutput=1

\usepackage{dsfont}
\usepackage{amsmath,amssymb,amscd,amsfonts,mathtools}
\usepackage[utf8]{inputenc}
\usepackage[T1]{fontenc}

\newtheorem{thm}{Result}
\usepackage[toc,page]{appendix}
\usepackage{epsfig}
\usepackage{epstopdf}
\usepackage{latexsym}
\usepackage{graphicx}
\usepackage{subfig}
\usepackage{booktabs}
\usepackage{bbm}
\usepackage{color}
\usepackage{physics}
\usepackage{tensor}
\usepackage{tikz}
\usepackage{tcolorbox} 
\usetikzlibrary{matrix}
\usetikzlibrary{decorations.markings,calc,shapes,decorations.pathmorphing,patterns}
\usetikzlibrary{positioning}
\makeatletter
\def\@fpheader{\relax}
\makeatother

\author{Andreas Karch$^{a}$}
\author{Carlos Perez-Pardavila$^{a}$}
\author{Marcos Riojas$^{a}$}
\author{Merna Youssef$^{a}$}

\emailAdd{karcha@utexas.edu}
\emailAdd{cjp3247@utexas.edu}
\emailAdd{marcos.riojas@utexas.edu}
\emailAdd{myoussef@utexas.edu}

\affiliation{$^a$Weinberg Institute, Department of Physics, University of Texas, Austin, TX 78712, USA.}

\subheader{\begin{flushright}

\end{flushright}}

\title{Subregion Entropy for the Doubly-Holographic Global Black String}
\author{}
\date{}

\abstract
{We study the growth of entanglement entropy in a doubly holographic model of gravity for a spherical AdS black hole. Compared to previous work, which was limited to the case of planar black holes, this introduces an extra scale to the problem. This allows us to analyze the interplay between the reorganization of entanglement entropy due to island formation and the onset of the Hawking-Page phase transition and to find the appearance of a new critical black hole radius unrelated to the thermodynamics. We also find that the geometry of the Ryu-Takayanagi surface capturing the physics of islands exhibits drastically different behavior than in the planar case. 
}

\begin{document}
\maketitle

\section{Introduction}
\indent The AdS/CFT correspondence \cite{Maldacena:1997re,Gubser:1998bc,Witten:1998qj}, which is the most concrete realization of the holographic principle to date, is a conjectured duality between two seemingly different physical theories. One of its most remarkable aspects is that, since the correspondence relates a higher-dimensional theory of gravity in anti-de Sitter space (AdS) to a dual conformal field theory (CFT) on its boundary, it provides a consistent framework where the unitarity of a black hole's evaporation process can be studied. 

The crucial ingredient in recent calculations, which have yielded a unitary Page curve for the time-evolution of the entanglement entropy, has been entanglement islands \cite{Penington:2019npb,Almheiri:2019psf} -- regions of spacetime which, while seemingly disconnected from the holographic system, arise when the boundary region is coupled to a system with additional gravitational degrees of freedom. They are responsible for the late-time contribution to the entropy of a subregion on the boundary theory, and their location on a slice of AdS$_{d+1}$ is determined by a quantum extremal surface (QES) \cite{Engelhardt:2014gca} whose location, if perturbed, won't affect the measurements of entanglement entropy. 

\indent However, determining the location of these islands is highly nontrivial as one needs to have quantitative control of the quantum entanglement across the QES. But by introducing a boundary to the conformal field theory, one obtains a doubly holographic model where the entanglement entropy can be computed using the classical Ryu-Takayanagi (RT) prescription in one dimension higher \cite{Ryu:2006bv,Ryu:2006ef,Hubeny:2007xt,Faulkner:2013ana,Lewkowycz:2013nqa,Chen:2020uac,Chen:2020hmv}. In such models the boundary of the CFT, which we will refer to as the defect, is dual to a Karch-Randall (KR) brane where RT surfaces can connect, thus defining a quantum extremal surface and forming an island on the brane. More precisely, following previous work \cite{Geng:2021mic}, the island is defined as the region extending from the RT surface's anchor point on the brane to the other side of the thermofield double, and the family of anchor points on the brane to which RT surfaces can connect is defined as the atoll. RT surfaces that connect the brane to the boundary are called island surfaces, and when they connect the brane to the defect, the location where it attaches to the brane is known as a critical anchor.

One way to realize this model is by starting with a Karch-Randall (KR) brane on a $d$-dimensional slice of AdS$_{d+1}$. Through the AdS/CFT correspondence, this model has three equivalent descriptions \cite{Karch:2000ct,Randall:1999vf,Karch:2000gx,Takayanagi:2011zk,Fujita:2011fp} which we will refer to as the boundary, intermediate, and bulk pictures. The boundary picture is a unitary $d$-dimensional CFT ending on a $d-$1 dimensional defect, and the bulk picture is an AdS$_{d+1}$ spacetime with an embedded AdS$_d$ KR brane. We emphasize that the bulk picture is particularly notable because we can use this description to compute entanglement entropy using classical RT surfaces. The location of the RT surface in the bulk description determines the location of the QES in the intermediate picture. Here we have a $d$-dimensional CFT on the KR brane ending on the defect, with semi-transparent boundary conditions, allowing energy transfer between the KR brane and a non-gravitating bath housing a $d$-dimensional CFT. The intermediate description is useful because, once a Schwarzschild black hole is introduced on each AdS$_{d+1}$ spacetime slice, we can use it to interpret the areas of the bulk RT surfaces -- which determine the entanglement entropy of the radiation region in the non-gravitating bath -- as the entanglement entropy of the radiation escaping the black hole on the brane. 

We pause briefly to describe the coupling of the conformal field theory to a semi-transparent boundary, which makes sense from a physical perspective. There is a negative cosmological constant on the brane which necessitates the introduction of boundary conditions for the quantum fields. The conventional approach is to choose reflecting boundary conditions, and when this choice is made, AdS black holes essentially cannot evaporate on their own because the Hawking radiation they send to the boundary is reflected. One way to define an evaporation process for them is to couple the system to a heat bath \cite{Rocha} -- in this model, from the intermediate perspective, this is achieved through the transparent boundary conditions at the defect that separates the degrees of freedom on the brane from the degrees of freedom on the boundary CFT. 

The main difference between the setup we consider here, and the one previously explored in \cite{Geng:2020fxl, Geng:2021mic}, is that we are interested in spherical Schwarzschild black holes instead of planar ones. This means we use AdS in global coordinates. To understand the significance of this, let us first recall some facts about standard AdS/CFT at finite temperature without the brane. For global AdS, the topology of the boundary is $S^{d-1} \times S^1$, and that's where the CFT is located. If we instead place the CFT on flat background $R^{d-1}$, as has been done in previous work on double holographic islands \cite{Almheiri:2019hni,Almheiri:2019psy,Geng:2020qvw,Geng:2020fxl,Geng:2021mic,Geng:2021iyq}, this differs from a sphere $S^{d-1}$ in that the latter introduces an extra phase transition and associated instability. For a CFT on $R^{d-1}$, the temperature $T$ is the only scale, and therefore there is no notion of a hot or cold black hole; physics is equivalent at different temperatures. In this case, the boundary topology makes the AdS Schwarzschild black hole the only candidate solution in the bulk to express the thermal CFT on the boundary. As for the CFT on a sphere, we now have the radius of the sphere $R$ and temperature $T$ composing a dimensionless number $RT$  which can form nontrivial physics at different temperatures \cite{Witten:1998zw}. This is to say the CFT at finite $T$ on a sphere introduces two candidate solutions: the thermal AdS (empty AdS) and the Schwarzschild AdS black hole. It is the well-known Hawking-Page phase transition, above which the black hole is no longer the preferred phase, which signals that there exists a point at which a phase transition from a black hole to a thermal gas of gravitons occurs.

\begin{figure}
    \centering
    \includegraphics[width=\linewidth]{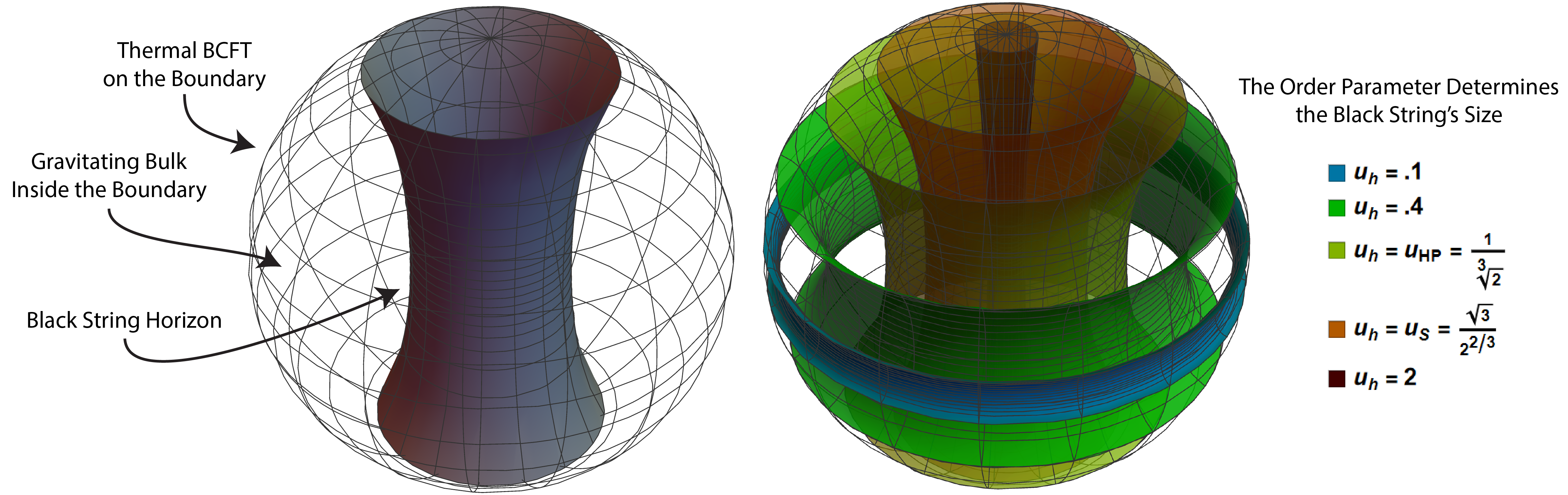}
    \caption{The diameter of the black string is determined by the order parameter in the theory, $u_h$. The bulk theory is dual to a thermal BCFT state on the surface of the sphere. 
    When $u_h$ becomes larger than  $u_{\text{HP}} = \frac{1}{2^{1/3}}$  
    we encounter the Hawking-Page transition and the string is no longer the global minimum of the free energy.
    The black string reaches minimum temperature at the spinodal point $u_s = \frac{\sqrt{3}}{2^{2/3}}$, where the specific heat changes sign from positive to negative. Also see Figure \ref{fig:bhtemperature}.} 
    \label{fig:BS}
\end{figure}

We employ the black string solution, a well-known solution to vacuum gravity in AdS$_5$ or higher dimensional spacetimes, to realize this important phase transition of a finite temperature CFT on the sphere within double holography. The scale in the theory is represented by the parameter $u_h$, which determines the size of the Schwarzschild black hole on each spacetime slice -- see Figure \ref{fig:BS}. From the bulk perspective, introducing the black string in global AdS$_{d+1}$ is equivalent to placing spherical Schwarzschild black holes on each AdS$_{d+1}$ spacetime slice, including on the boundary on which the BCFT lives. From the intermediate perspective, this amounts to placing equal size black holes on the gravitating brane and the non-gravitating bath, which allows them to be in equilibrium while still exchanging Hawking quanta. From the boundary perspective, this is just a theory at finite temperature. In our case, the finite temperature in the non-gravitating bath is maintained by having a black hole in the bath as well. While somewhat unnatural from the field theory point of view, this situation serves just as well as a heat bath. It greatly simplifies the problem due to the underlying simplicity of the black string. Furthermore, since this is a static geometry, the RT surfaces can be constructed on a single time slice. 

\indent It can be seen in Figure \ref{fig:BS} that the black string has the topology of a cylinder. Such black string solutions are known to have an instability, corresponding to the rippling of the horizon, which is known as the Gregory Laflamme instability \cite{Gregory:1993vy}. This instability is first encountered for the global AdS black string when the black holes on each slice dip below a certain size. Early on it was believed that this instability is dual to the Hawking-Page phase transition of the braneworld black hole  \cite{Chamblin:2004vr,Gregory:2008br} based on early numerical evidence \cite{Hirayama:2001bi}. More recent numerical evidence \cite{Marolf:2019wkz} and analytic arguments in a large number of dimensions \cite{Emparan:2021ewh} have shown that the GL instability, in fact, only sets in for black holes smaller than the radius at the Hawking Page transition. This is in fact consistent with the Gubser Mitra conjecture \cite{Gubser:2000mm}, which states that mechanical instabilities in the higher dimensional bulk should be dual to local thermodynamic instabilities in the dual CFT. The Hawking Page transition being first order does not correspond to a local instability.

Our objective in this note is to examine whether the Hawking-Page transition encountered in the global AdS black string model at a certain temperature, set by the order parameter $u_h$, plays an important role in this doubly holographic model which places a BCFT on the surface of a sphere. We are interested in the interplay of islands with the Hawking-Page phase transition. In particular, we are interested in obtaining a connection between the Hawking-Page transition and a general phase transition that occurs for higher-dimensional RT surfaces at the critical angle \cite{Geng:2020fxl}. The area difference between the Hartman-Maldacena (HM) surface, which encodes the increasing entanglement entropy of the black hole on the brane, and the island surface is known to diverge to negative infinity at the critical angle when there is a black hole on the brane. However, note that the area difference vanishes there in empty AdS. But we are mainly interested in the fact that island RT surfaces do not exist at all below the critical angle in empty AdS, and that for any asymptotically AdS geometry, the entanglement entropy above the critical angle is governed by tiny islands: global minima that live in the asymptotic region near the defect. 

In this work, where we examine the RT phase structure for the global black string, we obtain new scale-dependent behavior realized at the brane's critical angle ($\theta_c$). We find there exists a hole in the atoll when the brane lies above $\theta_c$ -- namely, there is a region on the brane where RT surfaces cannot anchor. From the intermediate perspective, this leads directly to a discontinuous phase transition for quantum extremal surfaces at $\theta_c$. For fixed brane angles above the critical angle, the size of this region grows monotonically as the black hole decreases in size and vanishes at what we call the saturation angle, $\theta_s$. We also find that RT surfaces can connect to the boundary points both above and below the hole, which in some cases leads to multiple candidate RT surfaces that need to be compared. This gives multiple candidate RT surfaces that travel from the brane to the defect, and we have found that the critical anchors for these surfaces lie exactly at the boundaries of the hole -- see Figure \ref{fig:critical_anchor} and Figure \ref{fig:thehole}. The hole's origin lies in the non-monotonic behavior of $h(u)$, and some surprising connections to the photon sphere for the black hole on the brane will soon be outlined in a separate note. Note that, while the hole exists for any non-zero $u_h$, the size of the hole starts to become large on the order of the Hawking-Page phase transition. Finally, we remark that the area difference between the HM and island surfaces vanishes at $\theta_c$ for a special value of $u_h$ which we call $u_h^{\text{crit}}$ -- in the language of \cite{Geng:2020fxl}, the Page angle equals the critical angle for this value of the order parameter. It can be seen in Figure \ref{fig:phase_diagram} that, even though it is invisible to purely thermodynamic arguments, its value nonetheless plays an important role in the phase structure of this model.

\section{Setup and Equations of Motion}

\subsection{The Doubly-Holographic Global Black String}

This section provides an overview of the doubly-holographic black string model which we are using to model the entanglement entropy of an evaporating black hole on the KR brane. It is important to realize that there is a major difference between this setup and the one we explored in our previous note \cite{Geng:2021mic} -- in brief, the area density for global AdS$_{d}$ depends on the scale set by the size of the black hole. Once we place a spherical AdS$_d$ black hole on each AdS$_{d}$ spacetime slice, we will obtain the global AdS$_{d+1}$ black string.

We begin by writing down the metric for an AdS$_{d}$ black hole in a convenient coordinate system. Conventionally setting the AdS$_{d}$ curvature radius to $1$, one can write spherically symmetric metrics as:

\begin{equation}
ds^2 = \frac{1}{u^2}\left[-h(u) dt^2 + \frac{du^2}{h(u)} + d\Omega^2_{d-2}\right]\label{eq:spherical}.
\end{equation}

To obtain a spherical Schwarzschild black hole, which we will soon place on each constant angle $\mu$ slice, we choose the following form for $h(u)$:
\begin{equation}
h(u) = 1+u^2-\frac{u^{d-1}}{u_h^{d-1}}. 
\end{equation}
Decreasing the value of $u_h$ always increases the size of the black hole. The coordinates we are using differ from the more conventional ones -- the difference is that we made the coordinate substitution: 

\begin{equation}
r \rightarrow \frac{1}{u}.
\label{eq:radial}
\end{equation}
and defined:
\begin{equation}
h(u) \equiv u^2 f\left(\frac{1}{u}\right),
\end{equation}
where $f(1/u)=f(r)$ is the standard blackening function:

\begin{equation}
f(r) = 1 + r^2 - \frac{\omega_{d-1} M}{\ r^{d-3}}.
\end{equation}
Where $\omega$ is a constant that is introduced so that $M$ is the mass of the black hole. By comparing these two coordinate systems, we can see that the parameter $u_h$ is given by:

\begin{equation}
    u_h = \left(\frac{1}{\omega_{d-1} M}\right)^{\frac{1}{d-1}}
    \label{uhmass}
\end{equation}

The main advantage of using $u$ coordinates, instead of $r$ coordinates, is that the boundary of AdS lies at $u=0$ instead of at $r=\infty$; similarly, $u=\infty$ corresponds to $r=0$ in standard coordinates. From a numerical perspective, these coordinates are much easier to work with than the standard $f(r)$ coordinates, and they also helped us gain insight into the equations of motion. 

In order to make our setup doubly holographic, we take the standard approach of employing the black string by embedding a $d$-dimensional KR brane in AdS$_{d+1}$ \cite{Randall:1999vf,Karch:2000ct}. This can be done by adding the standard Randall-Sundrum (RS) term \cite{Randall:1999vf} to the Einstein action while restricting to subcritical brane tensions. The simplest approach is to treat this as an \textit{end-of-the-world brane} by orbifolding the original 2-sided RS geometry. For the black string geometry the brane is embedded in we use:

\begin{equation}
ds^2 = \frac{1}{u^2 \sin^2\mu}\left[-h(u) dt^2 + \frac{du^2}{h(u)} + d\Omega^2_{d-2} + u^2 d\mu^2\right]\label{eq:BSsphere}.
\end{equation}
In contrast to Poincare-patch AdS, where the transverse coordinates are Cartesian, the transverse coordinates form a sphere $S_{d-2}$ with a line element proportional to $d\Omega_{d-2}$. Due to the presence of the brane, the angular coordinate must be larger than the angle of the brane location as we are taking a positive subcritical brane tension. This removes part of the universe with the angle variable being limited to: 
\begin{equation}
     \mu \in [\theta_{b},\pi).
\end{equation}
In addition, The introduction of the KR brane induces an AdS radius on the brane, $b$, which differs from the bulk value, $l$, according to the following equation:
\begin{equation}
    b = \frac{l}{\sin \mu}.
    \label{eq:radiusbraneworld}
\end{equation}

\subsection{The Doubly-Holographic System and Instabilities}
Here we discuss the doubly-holographic system which we are studying, beginning with a review of the black string and some well-known instabilities: the Hawking-Page phase transition and the spinoidal point. 

\subsubsection{The Location of the Black String}

The location of the Schwarzschild black string in the bulk, which has been plotted numerically in Figure \ref{fig:BS}, can be determined by determining where the blackening function $\frac{h(u)}{u^2}$ vanishes, which is equivalent to solving for the roots of the \emph{non-monotonic} function $h(u)$:

\begin{equation}
    h(u) = 1 + u^2(\mu) - \frac{u^{d-1}}{u_h^{d-1}} = 0
\end{equation}
In our case, where $d=4$, we have:
\begin{equation}
    u_+ = \frac{1}{3} \left(u_h^3 + \left(\frac{2}{\chi}\right)^{\frac{1}{3}}u_h^5+ \left(\frac{2}{\chi}\right)^{-\frac{1}{3}} u_h\right),
    \label{eq:horizonscale}
\end{equation}
with $\chi\left(u_h\right)\equiv 2 u_h^6+3\left(9+\sqrt{81+12 u_h^6}\right)$ and $u_+ \in \mathbb{R}^{+}$, introduces a scale to the theory with $u_h$ acting as an order parameter. The advantage to focusing on $d=4$ is that the system has enough dimensions to have interesting dependence on scale. There are also some less interesting cases where $d=2$ or $d=3$, where the horizon is given by:

\begin{equation}
    R_2 = \frac{1}{2{u_h}}-\frac{\sqrt{1-4 {u_h}^2}}{2{u_h}}\text{ if }{u_h}<\frac{1}{2}
\end{equation}
\begin{equation}
    R_3 = \frac{{u_h}}{\sqrt{1-{u_h}^2}}\text{ if }{u_h}<1
\end{equation}
These systems lack interesting dependence on scale for different reasons -- the $d=2$ system is scale invariant, and the $d=3$ system is the BTZ black hole on the brane. Note that, for a general number of dimensions, the location of the horizon can be straightforwardly obtained using standard root-finding methods. 

\subsubsection{The Hawking-Page Phase Transition}

Before discussing the full braneworld geometry, let us focus on AdS$_d$ gravity and its CFT$_{d-1}$ dual. The standard lore is that conformal field theories, on a manifold $M$ of dimension $d-1$, can be studied holographically by summing over the contributions of different gravitational bulk theories, $B$, of dimension $d$, which have $M$ as their boundary. Our bulk theory $B$ is susceptible to an interesting phase transition, known as the Hawking-Page phase transition \cite{Hawking:1982dh}, because the thermal state on the boundary, $M$, of the bulk lives on a sphere. 

When the thermal boundary state lives on a sphere, there are two known bulk theories, $B$, which need to be compared \cite{Witten:1998zw} -- the first is empty AdS with a spherical boundary theory, also called thermal AdS,

\begin{equation}
    d s^2=\left(\frac{r^2}{b^2}+1\right) d t^2+\frac{d r^2}{\left(\frac{r^2}{b^2}\right)+1}+r^2 d \Omega^2,
\end{equation}
and the second is a standard AdS Schwarzschild black hole, 
\begin{equation}
    d s^2=\left(\frac{r^2}{b^2}+1-\frac{w_{d-1} M}{r^{d-3}}\right) d t^2+\frac{d r^2}{\left(\frac{r^2}{b^2}+1-\frac{w_{d-1} M}{r^{d-3}}\right)}+r^2 d \Omega^2.
    \label{eq:bhmetric}
\end{equation}
Here $\omega_{d-1}$ is a constant we introduce so that $M$ is the mass of the black hole, but its precise value does not play an important role in our discussion. 

The difference between the infinite actions associated to these geometries, $I$, was first computed for three spatial dimensions, $d-1=3$, by Hawking and Page \cite{Hawking:1982dh}. It is also computed by Witten \cite{Witten:1998zw} for a general number of spatial dimensions. The action turns out to be proportional to the volume and is given by:

\begin{equation}
    I=\frac{d-1}{8 \pi G_N} \lim _{R \rightarrow \infty}\left(V_2(R)-V_1(R)\right)=\frac{\operatorname{Vol}\left(\mathbf{S}^{d-2}\right)\left(b^2 r_{+}^{d-2}-r_{+}^{d}\right)}{4 G_N\left((d-1) r_{+}^2+(d-3) b^2\right)},
\end{equation}
where $r_+$ is the radius of the black hole in radial coordinates and $b$ is the $AdS_d$ curvature radius. These radial coordinates, $r$, are related to our coordinates, $u$, by the coordinate transformation\footnote{This differs from Equation \ref{eq:radial} because the $\sin(\mu)$ term arises when the braneworld Schwarzschild black hole is placed into a warped geometry.}:

\begin{equation}
    r = \frac{1}{u \sin \mu}
    \label{eq:conversion}
\end{equation}

The Hawking-Page transition occurs when $I$ changes sign -- in other words, an AdS black hole is unstable when its radius, measured from the origin \emph{on} the KR braneworld in $r$ coordinates, is less than the brane AdS radius:

\begin{equation}
    r_+ < b \implies \text{The AdS Black Hole on the Brane is Unstable} 
\end{equation}
The next step is to convert to the coordinates used in our note. Taking $u_+$ to be the location of the horizon in our coordinates, Equation \ref{eq:conversion} gives:

\begin{equation}
   u_+ \sin \mu = \frac{1}{r_+} > \frac{1}{b} = \frac{\sin \mu}{l}
\end{equation}
\begin{equation}
    u_+ > \frac{1}{l} \implies \text{The AdS Black Hole on the Brane is Unstable}
    \label{eq:unstable}
\end{equation}
Here we choose a bulk AdS radius $l=1$, so the Hawking-Page transition occurs when $u_+ = 1$. From Equation \eqref{eq:horizonscale}, it can be seen immediately that the Hawking-Page phase transition occurs at $u_h = 2^{-\frac{1}{3}}.$ Note that the change in the AdS radius on the brane was canceled by the $\sin(\mu)$ term from the warped geometry.

\subsubsection{The Spinodal Point}

\begin{figure}
    \centering
    \includegraphics[width=.7\linewidth]{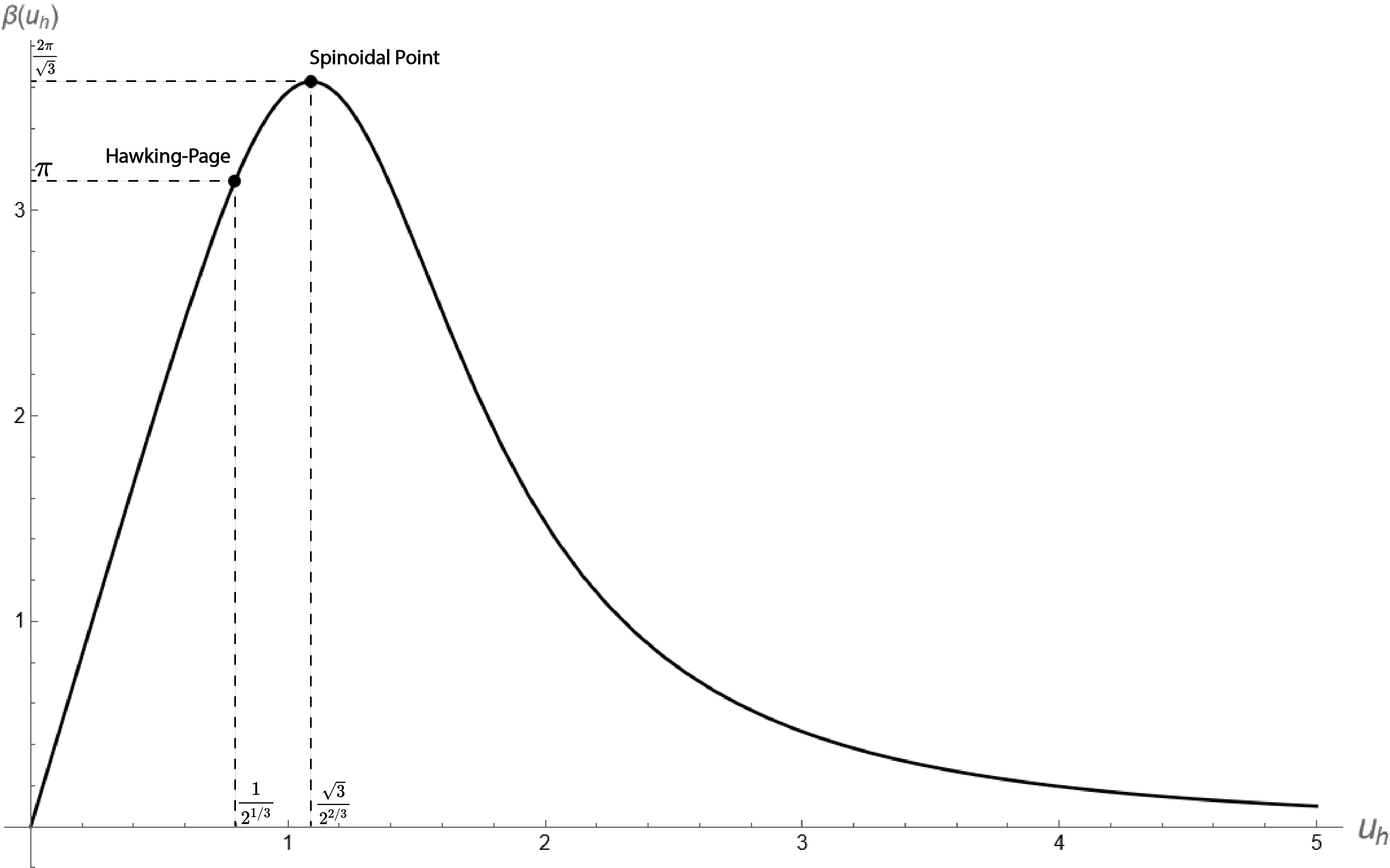}
    \caption{The temperature of the black hole scales non-monotonically with $u_h$, but for large black holes, the relationship is approximately linear. Unlike in asymptotically flat space, where black holes always have negative specific heat, AdS black holes have positive specific heat when they are large enough. The canonical ensemble in AdS is sensible because a temperature can be defined in the dual theory. For $\beta$ above the Hawking-Page phase transition, where the radius of the black hole is the AdS radius, the system will transition to a gas of thermal radiation. Such black holes cannot be in thermal equilibrium in the canonical ensemble. For $u_h$ past the spinodal point, black holes become hotter as they evaporate, so their specific heat is negative. Since AdS black holes do not exist above the value of $\beta$ where the spinodal point occurs, they do not contribute at all to the canonical ensemble.}
    \label{fig:bhtemperature}
\end{figure}

To facilitate our review of the spinoidal transition, we derive a well-known and generically useful result that relates the temperature of a black hole to the derivative of its blackening function at the horizon. Following the standard procedure \cite{Witten:1998zw}, we perform a Taylor expansion in the near horizon region:

\begin{equation}
ds^2 =  \frac{dr^2}{f(r)}+f(r) dt^2.
\end{equation}

When working in Euclidean signature, the metric encounters a conical singularity at the horizon which poses a problem for the periodic time coordinate. This can be repaired by carefully tuning the circle's circumference and amounts to fixing an inverse temperature, $\beta$, which makes the time-circle smooth at and outside the horizon. Start by noting that the metric near the horizon, written in Euclidean signature, is asymptotically flat with the periodic $\phi$ playing the role of a time coordinate:

\begin{equation}
ds^2 =  dr^2 + r^2 d\phi^2 
\end{equation}
Defining a new coordinate that is measured from the horizon, $\xi$, the Taylor expansion for the metric near the horizon yields: 

\begin{equation}
f(\xi) = f(r_+)+f’(r_+)(\xi-r_+)+ \mathcal{O}(\xi^2)
\end{equation}
One can match coefficients in the region near the horizon while pairing the periodic coordinates -- that is, one should match the time coordinate, $t$, with the $\phi$ coordinate. Solving the resulting system of equations gives an expression for the differential coordinate along the circle, $d\phi$:
\begin{equation}
\frac{d\xi^2}{f’(r_+)\xi}+f’(r_+)\xi dt^2 = d\rho^2 + \rho^2 d\phi^2 
\end{equation}
\begin{equation}
d\phi = \frac{f’(r_+)}{2}dt 
\end{equation}
When working in Euclidean signature, the system's time coordinate is periodic in the system's inverse temperature, $\beta$. Requiring that the time coordinate makes it "all the way around" the circle at the horizon yields the known expression for the inverse temperature of the system: 
\begin{equation}
    2 \pi = \frac{f'(r_+)}{2} \beta \implies \beta = \frac{4\pi}{f'(r)\rvert_{r_+}}
\end{equation}
The blackening function for the spherical AdS Schwarzschild black hole, given by Equation \eqref{eq:bhmetric}, then gives an expression for its temperature: 
\begin{equation}
\beta= \frac{4\pi}{f'(r)|_{r_+}} = \frac{4\pi r_+b^2}{(d-1)r_+^2 + (d-3)b^2} = \frac{4 \pi   u_+b^2}{(d-1)+ (d-3) b^2 u_+^2}
\end{equation}
The local minimum occurs when the derivative with respect to $r_+$ vanishes:

\begin{equation}
r_+ = b\sqrt{\frac{d-3}{d-1}} \implies r_+ = \frac{b}{\sqrt{3}} \iff u_+ = \sqrt{3}
\end{equation}
As in the previous section, the relationship \ref{eq:radiusbraneworld} between the AdS radius on the brane, $b$, and the AdS radius in the bulk, $l$, cancels the $\sin(\mu)$ term that arises due to the warped geometry. It can be seen in Figure \ref{fig:bhtemperature} that, at least for large black holes, $u_h$ is roughly proportional to the inverse temperature. This can be shown analytically. In the limit of large $M$, we can use an equation from \cite{Witten:1998zw}, together with Equation \ref{uhmass}, to show that the temperature of the black hole is approximately: 

\begin{equation}
\beta = \frac{4\pi}{f'(r)|_{r_+}} \approx \frac{4 \pi b^2}{(d-1)(\omega_{d-1} b^2)^{\frac{1}{d-1}} (M)^{\frac{1}{d-1}}} = u_h \left(\frac{4\pi }{d-1}\right)b^{\frac{2(d-2)}{d-1}}.
\end{equation}

\subsubsection{The Gregory Laflamme Instability}

From the perspective of the intermediate picture, where the conformal boundary and the KR brane are connected through the defect, the black string is perceived as two eternal AdS Schwarzschild black holes. These correspond to where the black string crosses the gravitating KR-brane and the non-gravitating conformal boundary, and one black hole lives in each region. Since the brane is gravitating, the AdS-Schwarzschild black hole on the brane will be susceptible to the Hawking-Page phase transition. 

Beyond a certain value of $u_h$, the cylindrical topology of the black string is unstable to small perturbations which lead to the formation of a series of black holes along the extra dimension \cite{Gregory:1993vy}. This is the well-known Gregory Laflamme instability for the black string. 
Recent results on this instability \cite{Marolf:2019wkz,Emparan:2021ewh} indicate that the Gregory Laflamme instability in fact happens at black hole radii even smaller than the spinodal point. This is rather surprising as one would have thought that the local thermodynamic instability that sets on at the spinodal should be represented by exactly such an instability of the black string against small fluctuations. It would be interesting to understand the interplay between Gregory Laflamme and the well-know thermodynamic transitions of the black hole on the brane in more detail. In this work we will not consider Gregory Laflamme any further and will focus on the interplay between Hawking Page, the spinodal point, and the formation of islands.

\subsection{Enforcing The Ryu-Takayanagi Prescription}

Now that we have constructed our doubly-holographic model, our objective is to determine the time-dependent entanglement entropy between the two black holes in the intermediate picture by computing the areas of classical RT surfaces anchored to the defect in the bulk picture. To do this we will need to follow the RT prescription.

The RT prescription, which requires us to determine the minimal extremal surfaces $\Sigma$ that are homologous to $\mathcal{R}$ \cite{Fujita:2011fp,Almheiri:2019yqk}, leads directly to the formation of islands when these surfaces terminate on the end-of-the-world brane \cite{Penington:2019npb,Almheiri:2019psf}. First we consider a generic interval on the brane $\mathcal{I}$---a ``candidate'' island. We then determine extremal surfaces $\Sigma$ that satisfy the homology constraint,
\begin{equation}
\partial\Sigma = \partial \mathcal{R} \cup \partial \mathcal{I}.
\label{eq:homology_constraint}
\end{equation}
The next step is to apply a variational principle to the \textit{area density functional} $\mathcal{A}$, which can be obtained by computing the square root of the determinant of the constant-time slices of the metric. This determines the area of the surface up to a factor, proportional to the volume of the suppressed dimensions, that does not affect the phase structure:
\begin{equation}
\mathcal{A} = \int \frac{d\mu}{(u\sin\mu)^{d-1}} \sqrt{u^2 + \frac{u'(\mu)^2}{h(u)}}, 
\label{eq:actionu}
\end{equation}
where $h(u)$ is given by: 

\begin{equation}
    h(u) = 1 + u^2(\mu) -\frac{u^{d-1}}{u_h^{d-1}}.
\end{equation}
The action parameterized as $\mu(u)$ can be obtained using the same procedure:

\begin{equation}
    \mathcal{A} = \int \frac{du}{(u \sin\mu)^{d-1}} \sqrt{u^2 \mu'(u)^2 + \frac{1}{h(u)}}
    \label{eq:actionmu}
\end{equation}

Now we vary the area functional to determine the $\Sigma$ and corresponding $\mathcal{I}$ for which the area is minimized. Note that, in contrast to our previous work, the size of the black hole plays an important role in this note. While we can bring the $u_h$ outside the integral for the action as an overall prefactor $u_h^{2-d}$  when the black string is large, we cannot make this simplification for a scale-dependent black string in global AdS.\footnote{For this reason, for ease of comparison with our previous note we have rescaled the axis by $u_+^2$ in Figure \ref{fig:Areas}.} In our case, the size of the black hole -- which now depends on $u_h$ -- introduces a scale to the theory which determines the width of the black string and plays a major role in the phase structure.

The Euler-Lagrange equations for the action \eqref{eq:actionu} will be ordinary differential equations because of the parameterizations $u = u(\mu)$ and $\mu = \mu(u)$.  As discussed in \cite{Geng:2020fxl}, the boundary terms in the variation of $\mathcal{A}$ vanish by imposing boundary conditions on $\Sigma$. Following the same argument, we impose a \textit{Dirichlet condition} at the conformal boundary and a \textit{Neumann condition} -- requiring that the first derivative $u'$ vanishes -- at the brane. The Neumann boundary condition demands that RT surfaces anchor to the brane at right angles. Depending on which coordinate system -- $u(\mu)$ or $\mu(u)$ -- we choose to use, we have:

\begin{equation}
    u'(\theta_b) = 0
    \label{eq:BCu}
\end{equation}

\begin{equation}
    \mu'(u_b) = \infty
    \label{eq:BCmu}
\end{equation}

It's possible to gain some physical intuition into these boundary conditions. The first states that the RT surface is struck at a right angle at the brane, since increasing the angle does not change the radial distance. The second statement follows from taking the reciprocal.

\subsubsection{The Equations of Motion}

Since we wrote the action in terms of two different parametrizations, we can obtain the corresponding equation of motion for each of them separately. Varying the action \eqref{eq:actionu} and solving the resulting Euler-Lagrange equation for the second derivative $u''(\mu)$ yields:

\begin{equation}
\begin{array}{r}
u^{\prime \prime}(\mu)=-(d-2) h(u) u(\mu)+u^{\prime}(\mu)\left((d-1) \cot \mu-(d-3) \frac{u^{\prime}(\mu)}{u(\mu)}\right) \\
+\left(\frac{u^{\prime}(\mu)}{u(\mu)}\right)^2\left(\frac{u(\mu)^2 h^{\prime}(u)+2(d-1) \cot (\mu) u^{\prime}(\mu)}{2 h(u)}\right).
\end{array}
\label{eq:EOMu}
\end{equation}
We can similarly vary the action \eqref{eq:actionmu} and solve the resulting Euler-Lagrange equations for the second derivative $\mu''(u)$, which yields:

\begin{equation}
\begin{aligned} 
\mu^{\prime \prime}(u)=(d-2) h(u) u \mu^{\prime}(u)^3 +\left(\frac{\mu^{\prime}(u)}{u}\right)&\left((d-3)-(d-1) u \cot (\mu(u)) \mu^{\prime}(u)\right) \\ &-\frac{2(d-1) \cot (\mu(u))+u^2 h^{\prime}(u) \mu^{\prime}(u)}{2 u^2 h(u)} \end{aligned}
\label{eq:EOMmu}
\end{equation}

These equations of motion are useful in different situations. The first set of equations is typically more useful when solving the shooting problem from the brane, since the RT surface will lie at a right angle to the radial direction, which sets $u'(\mu)=0$. The second set is typically more useful when shooting from the defect, since the RT surface will point along the radial direction, which sets $\mu'(u)=0$. 

For example, when shooting from the brane we have $u'(\mu)=0$, which reduces our $u(\mu)$ equations of motion to: 

\begin{equation}
u''(\mu) = -(d-2) h(u) u(\mu)
\end{equation}
It follows immediately that RT surfaces that shoot from the horizon of the black hole, where $h(u)=0$, will remain on the horizon until they reach the boundary. Indeed, in such cases we have $u''(\mu)=0$ and $u'(\mu)=0$, so $u(\mu)$ lies at a stationary point.

Similarly, when shooting from the defect we have $\mu'(u)=0$, which reduces our $u(\mu)$ equations of motion to:

\begin{equation}
\begin{aligned} 
\mu^{\prime \prime}(u)= -\frac{(d-1) \cot (\mu(u))}{ u^2 h(u)} \end{aligned}
\label{EOMmu}
\end{equation}
Under similar reasoning, this equation is stationary when $\mu=\pi/2$. Such solutions correspond to the Hartman-Maldacena surface \cite{Hartman:2013qma} which travels from the defect, crosses the black hole horizon, and travels through the Einstein-Rosen bridge to the other side of the thermofield double.

\section{Numerical Results}

Here we present our main results. Our objective is to understand how the scale-dependent location of the horizon, given by Equation \ref{eq:horizonscale}, affects the phase structure, especially at the critical angle. We find that introducing a scale induces a discontinuous phase transition in the entanglement wedge at the critical angle for the brane. This is in contrast to \cite{Geng:2020fxl}, where the scale-less theory had a continuous phase transition at the critical angle. We will explain how this new behavior comes about due to a "hole" in the \emph{atoll}, which is the region on the brane where RT surfaces can shoot from the brane to the bath. 

The standard approach is followed throughout. By using our equations of motion, given by Equations \ref{eq:EOMu} and \ref{eq:EOMmu} respectively, together with the appropriate boundary conditions, given by Equations \ref{eq:BCu} and \ref{eq:BCmu}, we identify the candidate extremal surfaces which compete to dominate the entropy functional, given by Equations \ref{eq:actionu} and \ref{eq:actionmu}. As in our previous work \cite{Geng:2020fxl}, we are interested in the boundary conditions satisfying the homology constraint in Equation \ref{eq:homology_constraint}, namely, $\Gamma=0$, which anchors the RT surfaces to the defect and determines which regions can, in principle, be reconstructed using information localized at the defect.

\subsection{Numerical Approach: The Shooting Method}

The RT surfaces were obtained using the following procedure, which amounts to using the shooting method at the defect: 

\begin{enumerate}
  \item Using the equations of motion -- in our case, Equation \ref{eq:actionmu} --  expand $\mu(u)$ in series in the asymptotic region near the defect. Solutions can then be obtained near the defect using an asymptotic expansion.
  \item The asymptotic expansion cannot satisfy the boundary conditions, given by Equation \ref{eq:BCmu}, which require $\mu'(u)$ to diverge at some finite value of $u$. One must define a suitable cutoff region for the asymptotics and switch to numerics when the asymptotic solution crosses the boundary of that region. Since the equations of motion are second order, the values of $\mu(u)$ and $\mu'(u)$ at the cutoff region will suffice to perform the numerics. 
  \item For some given parameters, the first derivative $\mu'(u)$ is found, numerically, to diverge to positive infinity at some angle $\theta_b$. Therefore the boundary conditions, given by Equation \ref{eq:BCmu}, are obeyed at this angle. While such  solutions will not necessarily be unique, each corresponds to an RT surface anchored to a KR brane at an angle $\theta_b$.
\end{enumerate}
This method should be widely applicable to doubly-holographic models where RT surfaces anchored to the defect need to be calculated.

\subsection{The Series Expansion}

The series expansion near the defect for $\mu(u)$, which is needed to perform the shooting method described in the previous section, can be written as:

\begin{equation} \mu(u) = \mu(0) + \alpha_1 \mu'(0) + \alpha_2 \mu''(0) + \alpha_3 \mu'''(0) + \dots 
\label{general_museries}
\end{equation}
In our coordinate system the RT surface intersects the defect at a right angle at $\mu=\frac{\pi}{2}$, so we have $\mu(0)=\pi/2$ and $\mu'(0)=0$. It is straightforward to show that $\mu''(0)=0$ by using the equations of motion for $\mu(u)$ in \eqref{EOMmu}, and so: 

\begin{equation} \mu(u) = \frac{\pi}{2} + \alpha_3 \mu'''(0) + \dots 
\label{museries}
\end{equation}
This equation gives $\mu(u)$ in the asymptotic region as a function of just one parameter $\alpha_3$, with each of its values potentially yielding a solution satisfying the boundary conditions for a brane at some angle $\theta_b$. Result 1, see Appendix A, points out that the RT surfaces anchored to the defect will not necessarily be unique, and in fact, we have obtained multiple solutions for each brane angle above the critical angle. 

\subsection{Multiple Critical Anchors, and The Hole in the Atoll}

\begin{figure}
    \centering
    \includegraphics[width=\linewidth]{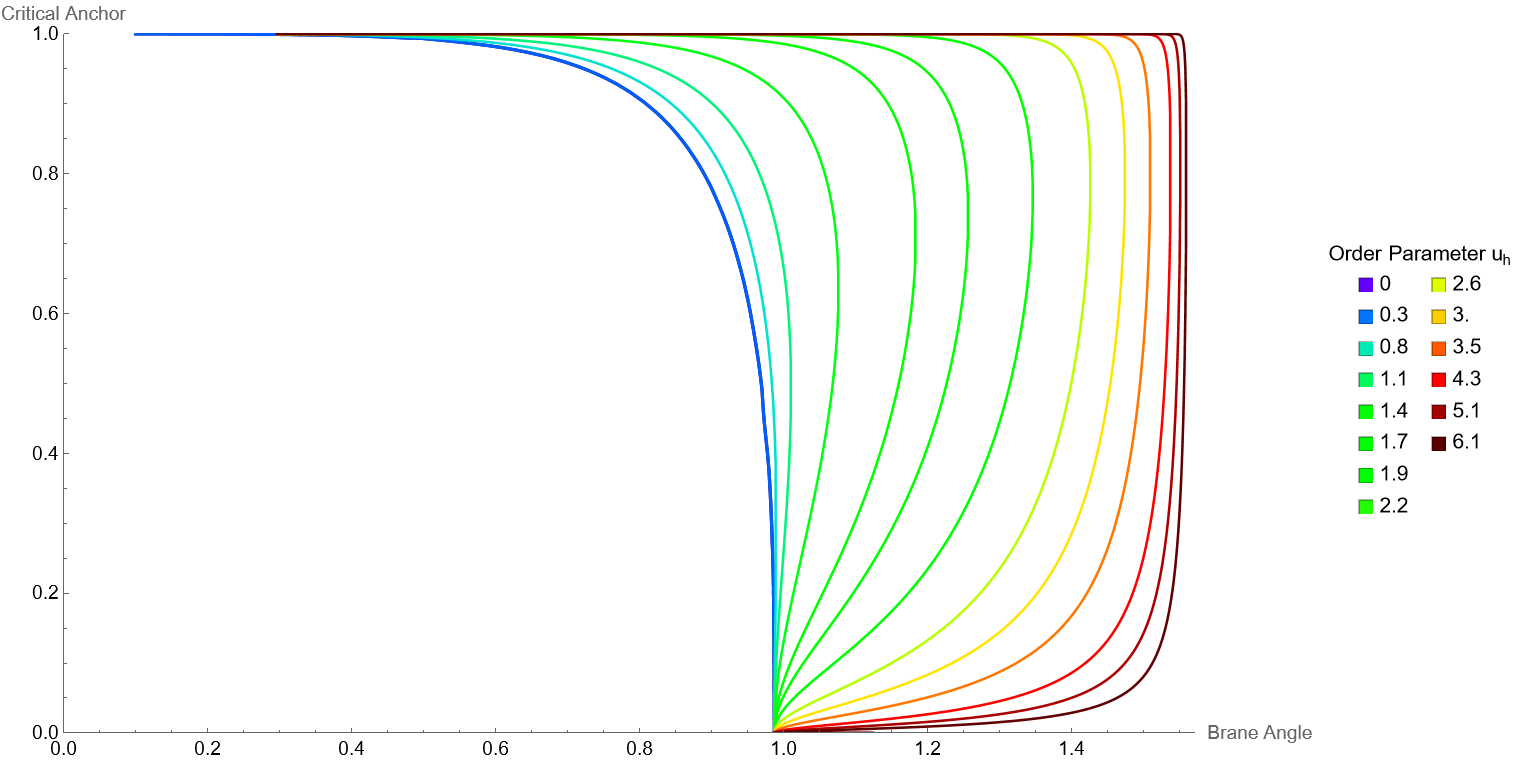}
    \caption{This figure shows the critical anchor on the brane, as a function of brane angle, as a percentage of the horizon radius, for various choices of order parameter $u_h$. The black string's diameter, and correspondingly, the black hole's size on the KR brane,  decreases (increases) as $u_h$ is increased (decreased). The Hawking-Page transition lies at $u_h = \frac{1}{2^{1/3}} \approx .8$. We recover our result for a large black string, discussed in \cite{Geng:2020fxl}, by taking the limit  $u_h \rightarrow 0$. There is a "hole" on the atoll when the critical anchor plot is not 1-to-1, which occurs for any non-zero value of $u_h$. As $u_h \rightarrow \infty$, the black string's diameter shrinks to a point, and the plot approaches a square shape. There is only one critical anchor below the critical angle ($\theta_c \approx .98687$), but there are two critical anchors between the critical angle and the saturation angle  $\theta_s$.}
    \label{fig:critical_anchor}
\end{figure}

\begin{figure}
    \centering
    \includegraphics[width=\linewidth]{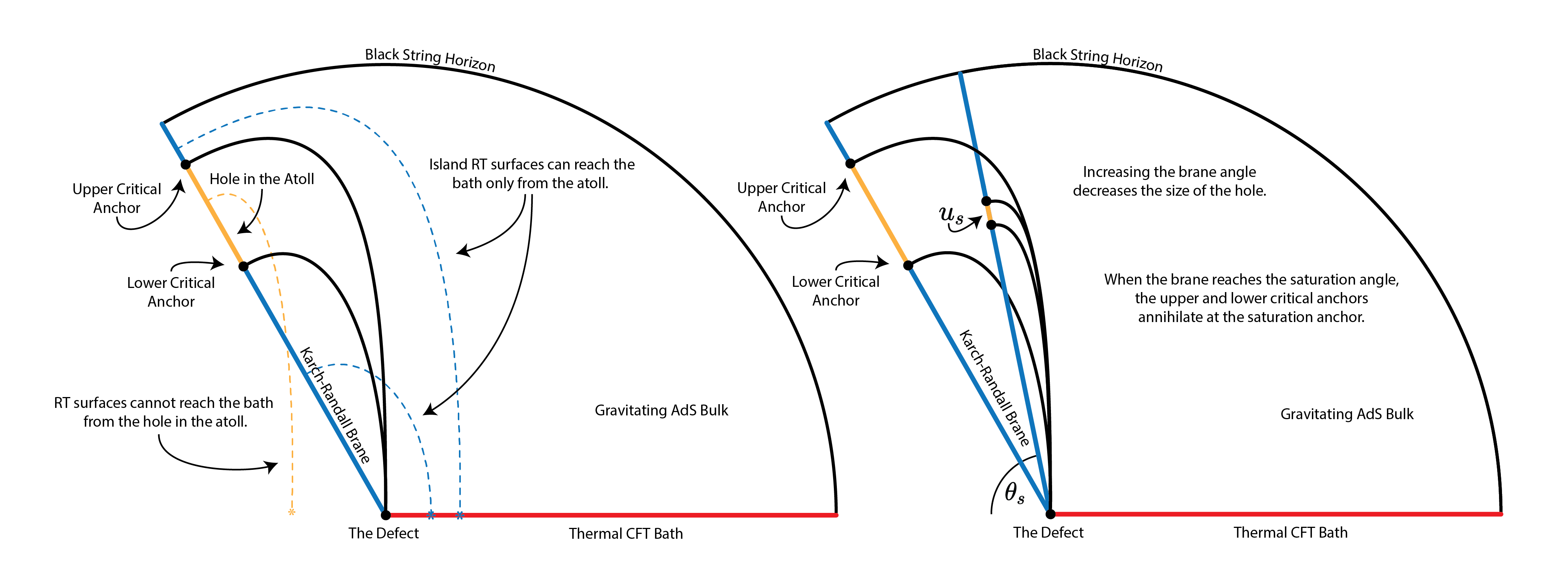}

    \caption{This cartoon illustrates typical RT surfaces, above the critical angle $\theta_c$, for an eternal black string dual to a BCFT on the surface of a sphere. When the KR brane lies above the critical angle, RT surfaces can travel from the KR brane to the defect from either the upper or lower critical anchor. RT surfaces can reach the boundary only from the atoll, and the region between the critical anchors is always a hole in the atoll. Increasing the brane angle decreases the size of the hole; at the saturation angle, the upper and lower critical anchors annihilate at the saturation anchor. Note that similar observations apply to anchors within a finite distance of the defect on the non-gravitating bath. }
    \label{fig:thehole}
\end{figure}

It was shown in previous work \cite{Geng:2020fxl,Geng:2021mic} that RT surfaces are restricted to parts of the brane in higher dimensional models. The region where candidate RT surfaces can anchor to the KR brane, thereby forming islands on the brane when they dominate the entropy functional, has been called the \emph{atoll}. The point where defect-anchored RT surfaces form an island by anchoring to the brane is called a \emph{critical anchor}. We can, for example, use these definitions to restate a relevant result from an earlier paper -- when there is a \emph{planar} black hole on each constant $\mu$ slice, as in \cite{Geng:2020fxl,Geng:2021mic}, the atoll is the region beyond the critical anchor on the brane. Note that, above the \emph{critical angle}, tiny island surfaces dominate the entropy functional. Please see Figure \ref{fig:tinyisland} and Appendix B for more details. 

Here we generalize our previous results by extending them to \emph{spherical Schwarzschild} black holes on each slice, with a horizon distance set by $u_h$. The scale set by $u_h$ leads to interesting new behavior -- note, however, that we can obtain our old results by taking the limit where $u_h \rightarrow 0$. 
We have found that -- above the critical angle -- there are \emph{two} critical anchors which border a region on the brane which we call the "hole" in the atoll. The critical anchors as a function of $u_h$ and the angle of the brane have been computed and can be seen in Figure \ref{fig:critical_anchor}. Those critical anchors are determined as a percentage of the horizon distance. The hole defines a region on the brane from which RT surfaces cannot reach the bath -- please see Figure \ref{fig:thehole} for an illustrative example.

The extent of the hole on the atoll depends on the value of $u_h$, according to Equation \eqref{eq:horizonscale}, with larger values of $u_h$ leading to smaller Schwarzschild black holes and larger holes in the atoll. Its size can be seen by inspecting Figure \ref{fig:critical_anchor} -- for any non-zero value of $u_h$, the critical anchor plot stops being 1-to-1 above the critical angle, $\theta_c$, and increasing the value of $u_h$ causes the plot to become not 1-to-1 within an increasingly wide strip of $\theta_b$ values. For those brane angles, there is more than one critical anchor, and the hole is defined as the region between the two critical anchors. 

\begin{figure}
    \centering
    \includegraphics[width=\linewidth]{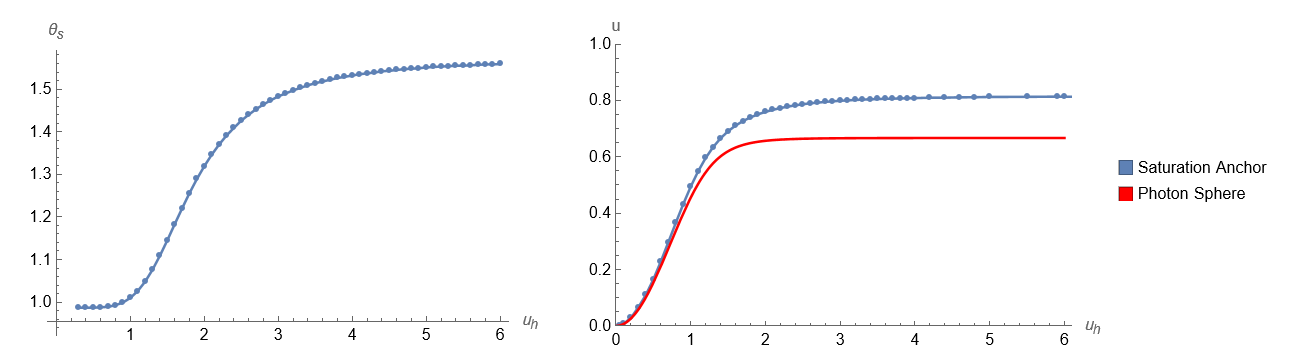}
    \caption{This figure shows the saturation angle for the brane on the left, and the saturation anchor for the brane on the right, both as a function of $u_h$. For a given value of $u_h$, the saturation angle is the largest brane angle with a hole on the atoll. As the brane angle increases, the hole shrinks to a point that we call the saturation anchor. Our numerics are consistent with Result \ref{thm:saturation_anchor}, in the Appendix, which states that the saturation anchor always lies inside the photon sphere.}
    \label{fig:saturation}
\end{figure}

It can be seen in Figure \ref{fig:critical_anchor} and Figure \ref{fig:thehole} that, for any fixed value of $u_h$, increasing the brane angle decreases the size of the hole up until what we call the \emph{saturation angle} $\theta_s$. The hole does not exist beyond that point because it shrinks to a point on the brane which we call the \emph{saturation anchor} $u_s$. Beyond the saturation angle, the atoll saturates the brane, and an RT surface can reach the bath from any point on the brane -- i.e., there is no hole in the atoll beyond the saturation angle. The saturation angles and saturation anchors are given in Figure \ref{fig:saturation}. We observe that in the limit where $u_h \rightarrow \infty$, the saturation angle approaches $\pi/2$, and the saturation anchor approaches $\approx 81\%$ of the horizon distance. This generalizes our result for the planar black string \cite{Geng:2020fxl,Geng:2021mic}, where the atoll first saturates the brane at the critical angle $\theta_c$ as we approach it from below and continues to saturate it above -- for the global black string, the atoll saturates the brane at the saturation angle $\theta_s > \theta_c$.

\subsection{Area Differences}

We have used the equations of motion to determine which extremal surfaces can reach the defect, located at $\Gamma=0$, and we have identified a "hole" in the atoll on the brane. There is a second question concerning these surfaces that is also interesting -- which bulk region is holographically dual to the defect, in the sense that we can use boundary information located on the defect to reconstruct its contents? This is equivalent to asking which extremal surface, for a given brane angle $\theta_b$, and KR braneworld Schwarzschild black hole size (set by $u_h$), dominates the entropy functional on a given time-slice.

The answer to this question depends critically on the tiny island surfaces, which always (never) dominate the area functional for $\Gamma=0$ above (below) the critical angle $\theta_c$. More details are available in \cite{Geng:2020fxl}, and we also review the tiny islands in Appendix B. To answer this question below the critical angle, where the tiny islands are never dominant, we have computed the area differences between the "candidate" RT surfaces: the Hartman-Maldacena surfaces, which plunge from the defect through the horizon, and the island RT surfaces, which travel from the defect to the brane. While these areas are formally infinite, their area difference is finite and given by: 

\begin{equation}
    \Delta A (t) = A_{IS} - A_{HM} (t)  \label{eq:DeltaA},
\end{equation}
for a fixed time slice, which we take to be $t=0$. Note that the Hartman-Maldacena surface is dominant (subdominant) when $\Delta A$ is positive (negative). In other words, when the area difference is positive, $A_{\text{IS}}>{A_\text{HM}}$ the Hartman-Maldacena surface is dominant on the initial time-slice and we get a non trivial Page curve. Similarly, when the area difference is negative, ${A_\text{HM}}>A_{\text{IS}}$ the island surface is dominant on the initial time-slice and we get a trivial Page curve.

This area difference depends on the time coordinate, $t$, because the Hartman-Maldacena (HM) surface traverses the Einstein-Rosen bridge -- hence, its area increases roughly linearly at late times, and its area will eventually exceed that of the island RT surface. So when the island RT surface does not start out dominant on the initial time slice, it will dominate after the area of the Hartman-Maldacena surface has grown by an amount equal to the area difference on the initial time slice. This happens at a coordinate time which we conventionally  call the Page time, denoted $t_p$. 

We are mainly interested in whether the islands dominate on the initial time slice. To determine which RT surface is dominant below the critical angle, it will suffice to compute $A(0)$, beginning with the HM surface. Since the anchor point on the bath lies on the defect (in other words, $\Gamma=0$) the HM surface drops straight down, on the $\mu=\pi/2$ slice, and crosses the black string horizon at $u=u_+$. Since $\mu'(u)=0$ along its trajectory, we can use  \ref{eq:actionmu} to show that its area outside the horizon is given by: 
\begin{equation}
    A_{HM} (0) = \int_{0}^{u_+} \frac{1}{u^3}\sqrt{\frac{1}{h(u)}} \label{eq:AHM}.
\end{equation}
This equation has closed-form solutions, which can be obtained using Mathematica but are cumbersome to write down. Meanwhile, the areas of the island RT surfaces must be obtained by solving the equations of motion \eqref{eq:EOMmu} using the shooting method, as described in Section 3.1, and evaluating the action \eqref{eq:actionmu} numerically. The results we obtained by carrying out this procedure can be seen Figure \ref{fig:Areas}. The next step, below, is to use these results to determine the phase structure.

\subsection{Phase Structure and the Critical Angle}

\begin{figure}
    \centering
    \includegraphics[width=.7\linewidth]{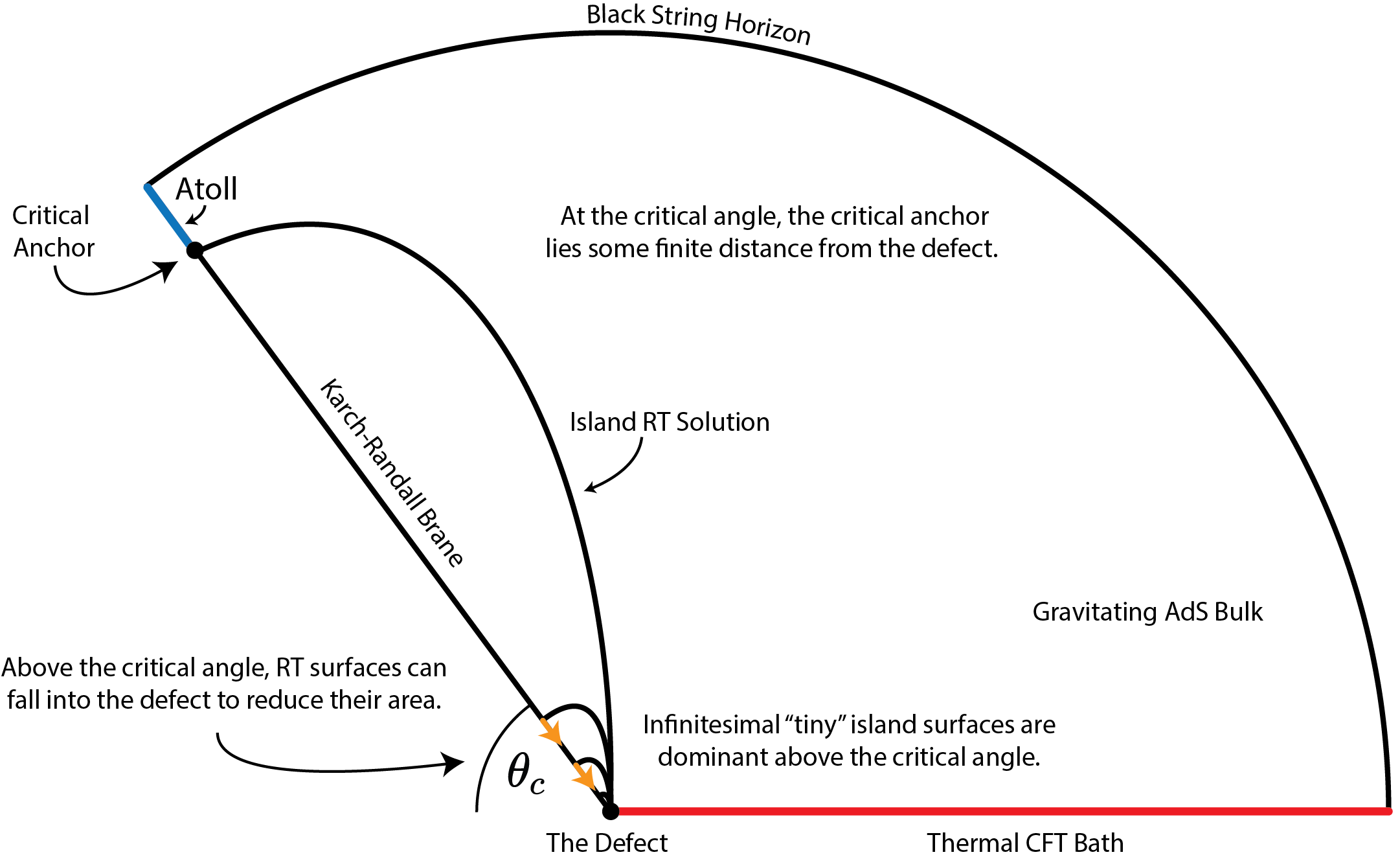}
    \caption{According to Figure \ref{fig:critical_anchor}, the critical anchor lies at a finite distance from the defect, which depends on the order parameter $u_h$, at the critical angle $\theta_c$. Then for $u_h \ne 0$, which corresponds to non-planar black holes on each slice, the phase transition at the critical angle is discontinuous because tiny island surfaces dominate the entropy functional at and above $\theta_c$. For a review of tiny islands, see the Appendix. By sending $u_h \rightarrow 0$, one recovers the result for planar black holes in our previous note \cite{Geng:2020fxl} -- the critical anchor drops into the defect, and the phase transition becomes continuous. Note that, in contrast to \cite{Geng:2020fxl}, the atoll does not saturate the brane at the critical angle when $u_h \ne 0$. }
    \label{fig:tinyisland}
\end{figure}

One of the key features of the phase structure for the black string is a discontinuous phase transition in the entanglement wedge at the critical angle, $\theta_c$, for the brane. It was established in \cite{Geng:2020fxl} that, since tiny island solutions always dominate the entropy functional above the critical angle, the island phase will dominate on the initial time slice above the critical angle. This leads to interesting new behavior for the black string -- when the angle for the brane, $\theta_b$, crosses the critical angle, $\theta_c$, the entanglement wedge shrinks discontinuously from finite to infinitesimal size -- see Figure \ref{fig:tinyisland} for an illustration. More details about the tiny islands can be found in \cite{Geng:2020fxl} and in the Appendix.

\begin{figure}
    \centering
    \includegraphics[width=\linewidth]{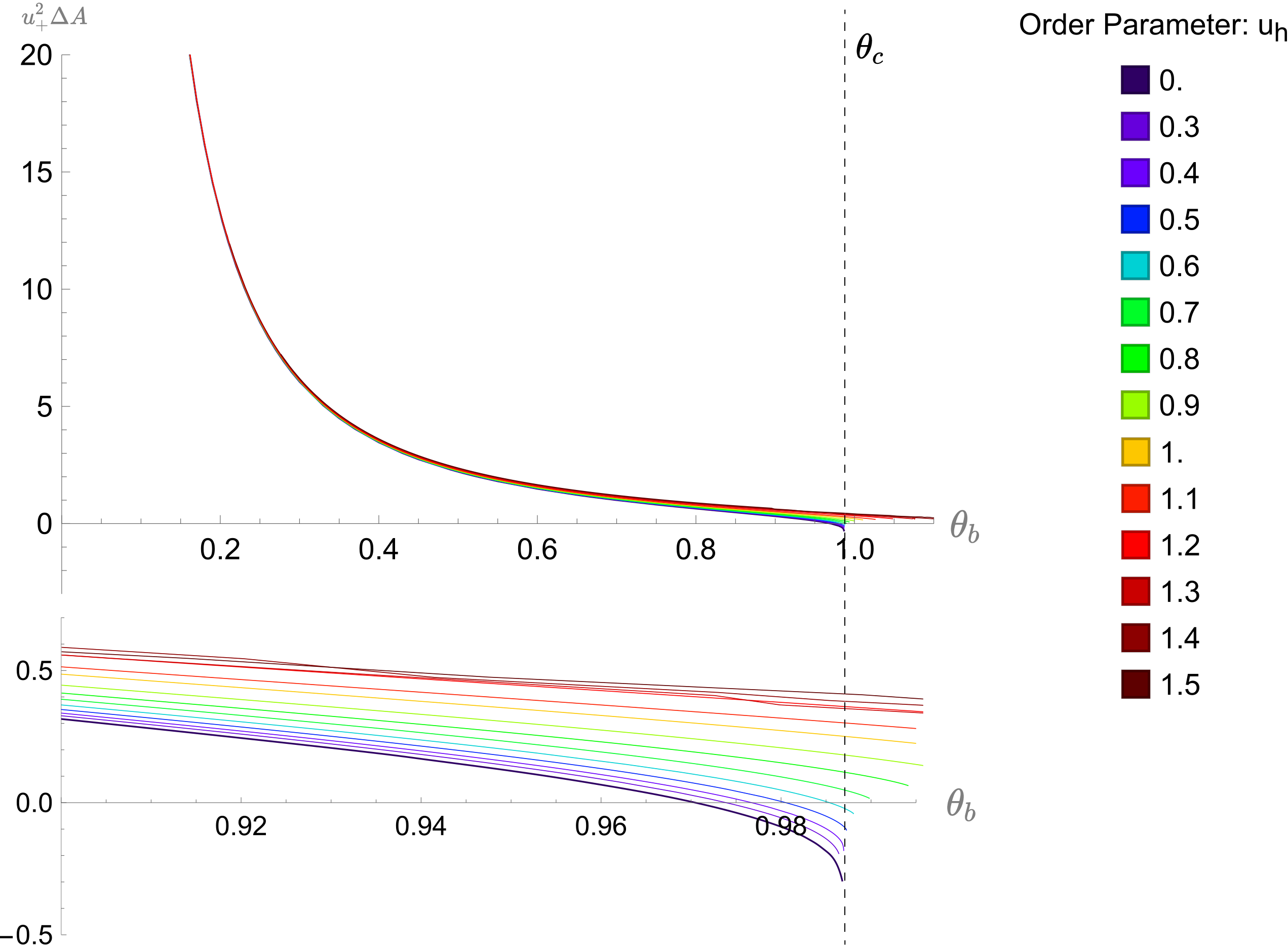}
    \caption{Here are the numerically computed area differences between the Hartman-Maldacena (HM) surface and the island RT surface for each value of $u_h$ below the critical angle $\theta_c \approx .98687$. The bottom panel zooms in on the region where the area difference vanishes. We find numerically that the Page angle equals the critical angle at roughly $u_h \approx .62777$. While the area differences \emph{above} the critical angle are also displayed for completeness, the actual dominant surface for $\theta > \theta_c$ will be the tiny island surface, for which the area difference actually diverges to negative infinity -- see the Appendix. By sending $u_h \rightarrow 0$, one obtains our old result \cite{Geng:2020fxl}, given by the thick dark curve, where the island RT surface degenerates into the tiny island surface at the critical angle -- so for $u_h=0$, the area difference diverges to negative infinity at $\theta_c$. For ease of comparison with our previous note, the vertical axis has been rescaled by the square of the horizon radius $u_+$. }
    \label{fig:Areas}
\end{figure}

The area differences are also interesting and can be seen in Figure \ref{fig:Areas}, where our numerical data is presented as a function of brane angle, $\theta_b$, for various $u_h$ values.\footnote{The numerics become more challenging as $u_h$ increases.} Some general features from the planar model are preserved in the spherical scale-dependent case. For any fixed value of black hole size, determined by $u_h$ and Equation \eqref{eq:horizonscale}, the area difference decreases monotonically with the brane angle. Hence, as before, island surfaces trend toward dominance as the brane angle increases. On the other hand, the dependence on the scale, $u_h$, is new. We find that, for any fixed value of the brane angle, $\theta_b$, the area difference increases monotonically with $u_h$.\footnote{Since $u_+$ increases with $u_h$, the rescaling of our axis by $u_+^2$ does not affect this result.} According to \eqref{eq:horizonscale}, and Figure \ref{fig:BS}, increasing the value of $u_h$ decreases the size of the black hole on the KR brane and narrows the diameter of the black string. Hence, islands tend to lose dominance as the size of the black string decreases.

\begin{figure}
    \centering
    \includegraphics[width=0.8\linewidth]{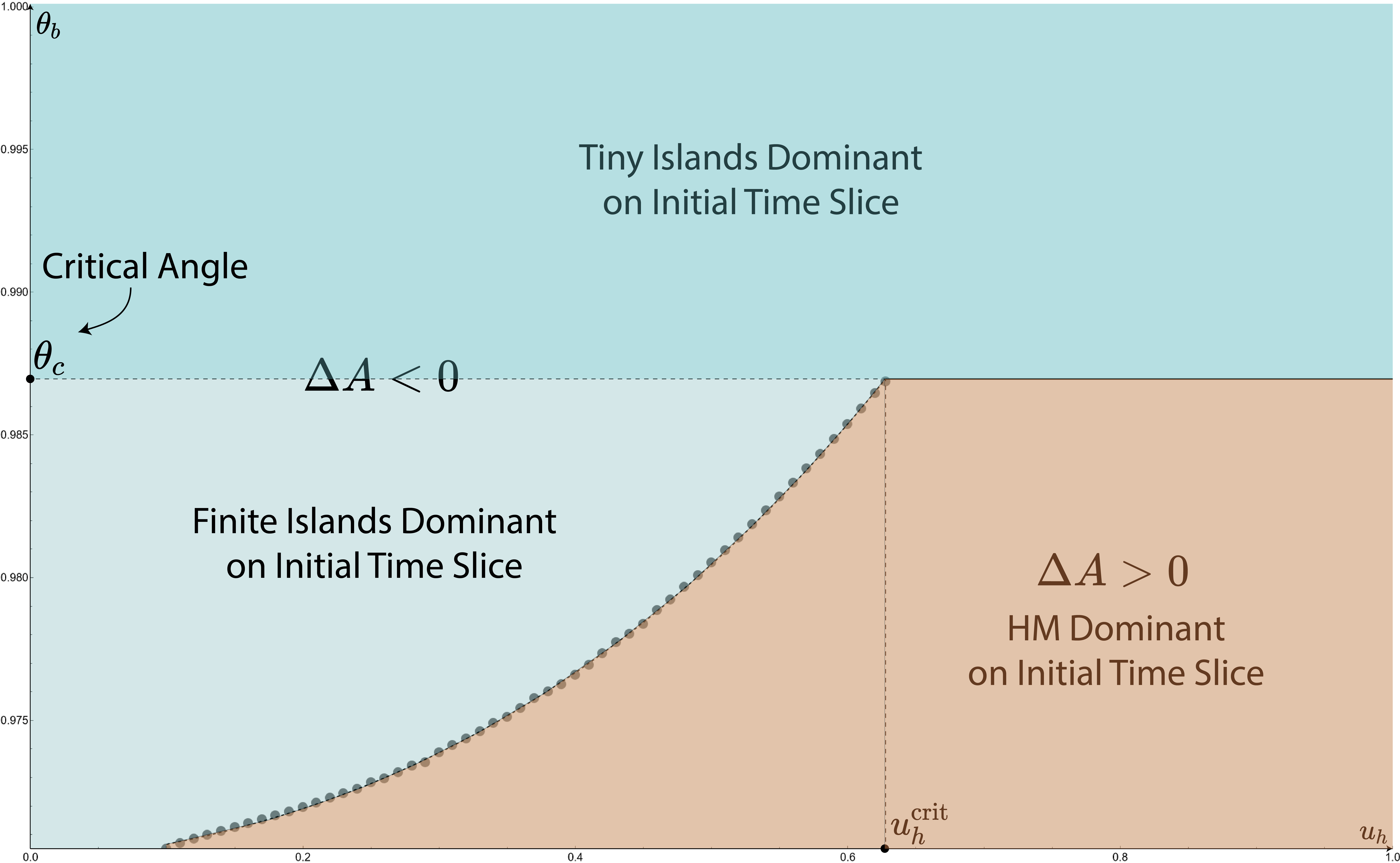}
    \caption{Here we present our phase diagram for the doubly-holographic global black string model. Since the thermal BCFT state lives on the surface of a sphere, there is a scale dependence in the model that depends on $u_h$ -- see Figure \ref{fig:BS} and Figure \ref{fig:bhtemperature}. Each point on this plot gives the brane angle for which the area difference vanishes, which we call the Page angle $\theta_p$. The Page angle was found to increase with $u_h$. The island surfaces are dominant when the area difference is negative. It was shown in previous work \cite{Geng:2020fxl} that the tiny island phase is dominant above the critical angle because the area difference diverges to $-\infty$. In contrast, finite-size islands dominate below the critical angle when the brane lies above the Page angle. When  $u_h > u_h^{\text{crit}}$, which is $\approx .62777$, the island surfaces are dominant only when the brane lies above the critical angle.}
    \label{fig:phase_diagram}
\end{figure}

The phase diagram for the eternal black string in $d=4$ is presented in Figure \ref{fig:phase_diagram}. The value of $\theta_b$ where the area difference vanishes is interesting since the island phase becomes dominant when the sign of the area difference changes. In previous work, we have called this value the Page angle $\theta_p$. Below the critical angle, $\theta_c$, there is genuine competition for dominance between the HM and island surfaces because the tiny islands are not dominant. For each value of $u_h$ below $u_h^{\text{crit}}$, which will be explained in a moment, we obtain a value of $\theta_b$ where the area difference vanishes below the critical angle. We find that the Page angle equals the critical angle at $u_h = u_h^{crit}$, where $u_h^{crit} \approx 0.62777$ -- beyond this point, the island phase is dominant precisely when  defect-anchored RT surfaces lie above the critical angle. Beyond this critical value for $u_h$, which amounts to fine-tuning the  radius of the Schwarzschild black hole on the brane, the entropy curve is constant if and only if the brane lies above the critical angle. 

Some additional connections to previous work are noted here. We have studied the case for $u_h=0$, for which the diameter of the black string is \emph{much} larger than the AdS radius, before \cite{Geng:2020fxl}. In that case, $\theta_p$ was slightly less than $\theta_c$ -- it was also found that, for empty AdS, $\theta_c$ was exactly equal to $\theta_p$ due to the scale invariance in the model. One of the authors has explored another situation where the Page and critical angles are equivalent under certain conditions after an Einstein-Hilbert term is used to introduce DGP gravity to the brane \cite{CarlosP}.

\section{Conclusion}

We have detailed the entanglement phase structure for a doubly-holographic black string formed from spherical AdS black holes placed on each spacetime slice. In such BCFT models, the boundary of the CFT, which we refer to as the defect throughout this note, is dual to a Karch-Randall (KR) brane where RT surfaces can connect. When they do so, this defines a quantum extremal surface (QES) and forms an island on the brane. The black holes on the brane and boundary are located where the black string crosses their respective regions, and we have computed the entanglement entropy between them by placing the anchor point for the bulk RT surfaces on the defect. It should be noted that since the RT surfaces in our model are homologous to the defect, our results apply equally well to the case where the bath is gravitating. 

Spherical AdS black holes differ from planar ones because they are susceptible to the Hawking-Page (HP) phase transition. This is connected to the fact that their dual thermal BCFT state lives on the surface of a sphere. Unlike in asymptotically flat space, where black holes have negative specific heat, AdS black holes below the HP phase transition can exist in stable thermal equilibrium with a heat bath. In contrast, when the inverse temperature, $\beta$, lies above the HP phase transition the system will transition to a gas of thermal radiation. This important phase transition 
is related to the observation that the blackening function $f(r)$ is not monotonic and the temperature achieves a minimum value at the spinoidal point. Some intriguing connections will be explored more generally in an upcoming note \cite{MarkR}.

We have shown that the phase structure, presented in Figure \ref{fig:phase_diagram}, for the entanglement entropy between two spherical AdS black holes exhibits drastically different behavior than the planar case explored in \cite{Geng:2020fxl,Geng:2021mic}. There is a direct connection to the diameter of the black string set by $u_h$, which also sets the system's temperature and acts as an order parameter in our model. Recall that when the island surface is dominant on the initial time slice, the entanglement entropy does not increase and there is a constant Page curve. For a fixed brane angle, as $u_h$ decreases, the area difference decreases. Thus islands become increasingly dominant as we decrease $u_h$. When the value of $u_h$ lies above a critical value, which we call $u_h^{\text{crit}}$, entanglement islands are always subdominant unless the KR brane lies above the critical angle $\theta_c$ identified in \cite{Geng:2020fxl}.

We have also found that the diameter of the black string in the bulk picture influences the existence of candidate RT surfaces. This is due to new scale-dependent behavior realized at the brane's critical angle. When the angle of the brane is greater than $\theta_c$, island RT surfaces are forbidden from anchoring to a region of the KR brane which we call the hole in the atoll. Our results are illustrated in Figures \ref{fig:critical_anchor} and Figure \ref{fig:thehole}. There are potentially up to two candidate RT surfaces that need to be considered for any choice of boundary anchor, with the extremal surfaces anchoring either above or below the hole. At the critical angle, the critical anchors -- defined in \cite{Geng:2020fxl} as the anchor points on the brane for extremal surfaces which connect the brane and the defect -- lie precisely at the boundaries of the hole. For fixed $u_h$, we find that the hole shrinks with increasing brane angle until it vanishes at what we call the saturation angle $\theta_s$. We also found that the critical anchors depend on the scale set by the diameter of the black string, and that the hole's existence induces a discontinuous change in the entanglement wedge at the critical angle. 

Several interesting connections exist between this work and our previous note \cite{Geng:2020fxl,Geng:2021mic}. In particular, we once again find that the area difference between the HM and island surfaces diverges to negative infinity at the critical angle -- the main difference is that the change in the size of the entanglement wedge is discontinuous instead of continuous. Note that, when we take the limit as $u_h \rightarrow 0$, we obtain the same results which were detailed in \cite{Geng:2020fxl,Geng:2021mic}. In particular, the hole in the atoll vanishes, the entanglement wedge again vanishes continuously at the critical angle, and the uniqueness of (finite) candidate island surfaces is restored. The restoration of the diverging area difference at the critical angle in this limit can be seen in Figure \ref{fig:Areas}. We also noted in \cite{Geng:2021mic} that the atoll need not be connected, and we have realized that in this particular doubly-holographic model. 

It would be interesting to analyze how the two competing surfaces, which appear for general $\Gamma$ on a non-gravitating bath when a scale is introduced, change the phase structure in \cite{Geng:2021mic}. Since tiny island solutions are unavailable when the RT surface anchors to a finite point on the boundary, there will be competition from the candidate RT surfaces anchoring above and below the hole on the atoll. We anticipate that, for fixed brane angles and anchor points on the boundary, there will be a phase transition between them at some value of $u_h$. We also expect that each boundary anchor will have its own saturation angle $\theta_s$. It might also be interesting to compute the Page time for various combinations of parameters in this model. 

As before, we conclude our discussion by noting that several general considerations should apply to a wide variety of braneworld models, even if the geometry is changed. First we note that, for doubly-holographic models with $d>2$, there will be a critical phase transition at the critical angle $\theta_c$. The existence of the hole on the atoll, which appears above the critical angle, is consistent with our predictions for the general structure of the atoll that were laid out in \cite{Geng:2021mic} -- the atoll should contain the region near the horizon on the brane, since the horizon is an extremal surface. It should also contain the region near the defect above the critical angle, since island surfaces exist in that region in empty AdS. Nonetheless, as we find here, it need not be connected. Our results are also consistent with our observation that increasing the brane angle should decrease the area difference between the HM and island surface. Physically one expects that the number of degrees of freedom on the defect decreases with angle, and the defect encodes the braneworld black hole; hence, increasing the brane angle should decrease the Page time. Finally, in this note, for the relatively small $u_h$ that we have studied, we have observed that the area difference increases with $u_h$. This suggests that smaller spherical AdS black holes take longer to saturate their entanglement entropy, at least in this system. It is possible that the area difference turns around as the black hole shrinks with $u_h$ and heats up. It would be interesting to study this observation in more detail.

\section*{Acknowledgments}

We'd like to thank Roberto Emparan for helping clarify the role of the Gregory Laflamme instability for the global black string which we mis-characterized in an earlier version of this note.
The work of AK, MR and MY was supported, in part, by the U.S. Department of Energy under Grant-No. DE-SC0022021 and a grant from the Simons Foundation (Grant 651440, AK). The work of CP was supported in part by the National Science Foundation under Grant No.~PHY-1914679 and by the Robert N. Little Fellowship.

\appendix 

\section{Useful Results about Islands and the Photon Sphere}

Here we quote some results from another paper, in preparation by one of the authors, which were relevant to the analysis of this note \cite{MarkR}.

\begin{thm}
For spherically symmetric blackening functions, there will be exactly one solution which travels from the brane to the defect \emph{unless} the black hole on the brane has a photon sphere. In such cases where there are multiple solutions, at most one solution will be outside the photon sphere. 
\label{thm:hole}
\end{thm}

\begin{thm}
The atoll is defined as the region on the brane where RT surfaces can anchor, forming an island. In some cases there is a "hole" in the atoll, with islands being possible on both sides of the hole. The hole exists only when there is a photon sphere; when it exists, part of the hole will always lie within the interior of the photon sphere. 
\label{thm:saturation_anchor}
\end{thm}

\section{Tiny Island Surfaces and Flat Page Curves}

A crucial role in the phase diagram is played by what we refer to as tiny island surfaces \cite{Geng:2020fxl}. These are tiny surfaces that connect the defect to the brane in the asymptotic region near the defect, and their dominance above the critical angle leads directly to the second-order phase transition noted in Section 3. The main point of this subsection is that tiny island surfaces are legitimate \textit{global} minima that dominate the area functional above the critical angle. Nonetheless, while tiny islands have many features in common with RT surfaces, we will review some important distinctions here. Note that the Page curve is flat when tiny islands are dominant, and furthermore, that these comments hold quite generally in higher dimensional AdS \cite{Geng:2020fxl}. 

Unlike tiny island surfaces, RT surfaces are local minima of the area functional. As such, their area functional vanishes under infinitesimal fluctuations of the RT surface. This is true both in the bulk and on the boundary -- indeed, the vanishing of the variation of the area with respect to the location of the boundary point is a physical requirement that gave rise to the condition \eqref{eq:BCu}. And since tiny islands do not satisfy condition \eqref{eq:BCu}, one might well wonder if these saddles are legitimate. 

But one must not lose sight of the fact that perfectly legitimate saddles of the area functional can be obtained under finite variations, so long as one obtains a \textit{global} minima for the area functional. Indeed, when the RT endpoints on the brane (the brane anchors) are allowed to fluctuate by a finite amount, the true global minima of the area functional can also occur at the boundaries of the range of the allowed anchors. And this is what tiny island surfaces are -- when the brane anchor approaches the defect under a finite variation, the corresponding RT surface degenerates into an infinitesimal surface localized very near the defect. 

It was found in \cite{Geng:2020fxl} that the regulated areas of these surfaces are always formally infinite, despite their tiny nature, but the infinity can be either positive or negative. Since these areas are \textit{positive} infinity below the critical angle, tiny islands are \textit{never} dominant in that regime. But it was also found that the areas are \textit{negative} infinity above the critical angle, which means they are \textit{always} dominant in that regime. The physical interpretation of this is as follows: when one remembers to include finite variations above the critical angle, one realizes that the would-be RT surface can simply slide toward the defect to decrease its area without bound. See Figure \ref{fig:tinyisland}. As its endpoint slides down the brane, the area difference between the island surface and the HM surface blows up to negative infinity -- in other words, the HM surface becomes infinitely larger -- and one obtains a tiny island that dominates the area functional. 

It may seem somewhat unsatisfactory to find that the area difference between these surfaces diverges, but this apparent infinity can be resolved using a limiting procedure. Consider a finite endpoint for the RT surface on the boundary, the boundary anchor, by sliding it slightly off the defect -- i.e., $\Gamma=\epsilon$ for some small $\epsilon$. One finds that the apparently infinite area difference is a large but finite one \cite{Geng:2020fxl}. To obtain the tiny island surfaces, one takes the limit as the endpoint on the boundary approaches the defect, i.e. $\epsilon \rightarrow 0$. The area difference diverges to negative infinity and one realizes that, for $\Gamma=0$, the tiny islands are indeed the dominant RT surfaces above the critical angle. In such cases, the tiny islands are dominant on the initial time slice and one obtains a flat Page curve.

\newpage

\bibliographystyle{JHEP}
\bibliography{references}
\end{document}